\documentclass[apjl]{emulateapj}
\usepackage{times}
\usepackage{amssymb}
\usepackage{epsfig}

\usepackage{graphicx,graphics}
\usepackage{dcolumn}
\usepackage{natbib}
\usepackage{times}
\def\kmm#1  {{\bf [KMM:~ #1]~}}
\def\new#1 {{\bf #1 }}
\def\cut#1 {\sout{#1} }

\newcommand{\hi}{H{\sc i}}

\newcommand{\beq}{\begin{equation}}
\newcommand{\eeq}{\end{equation}}

\newcommand{\kms}{km~s$^{-1}$}
\newcommand{\ts}{{\rm T}_s}
\newcommand{\tk}{{\rm T}_k}
\newcommand{\nhi}{{\rm N_{\sc HI}}}
\newcommand{\nlim}{{\rm N_{\sc LIM}}}
\newcommand{\cm}{cm$^{-2}$}

\shorttitle{An H{\sc i} column density threshold for CNM formation}
\shortauthors{Kanekar et al.}

\begin{document}

\title{An H{\sc i} column density threshold for cold gas formation in the Galaxy}

\author{N. Kanekar\altaffilmark{1},
R. Braun\altaffilmark{2},
N. Roy\altaffilmark{3}
}

\altaffiltext{1}{Ramanujan Fellow, National Centre for Radio Astrophysics, TIFR, 
Ganeshkhind, Pune-411007, India; nkanekar@ncra.tifr.res.in}
\altaffiltext{2}{CSIRO Astronomy and Space Science, Epping, NSW 1710, Australia}
\altaffiltext{3}{Jansky Fellow, National Radio Astronomy Observatory, 1003 Lopezville Road, Socorro, NM87801, USA}

\begin{abstract}

We report the discovery of a threshold in the H{\sc i} column density
of Galactic gas clouds below which the formation of the cold phase of
H{\sc i} is inhibited. This threshold is at ${\rm N_{\sc HI}} = 2
\times 10^{20}$~per~cm$^{2}$; sightlines with lower \hi\ column
densities have high spin temperatures (median ${\rm T}_s \sim 1800$~K), 
indicating low fractions of the cold neutral medium (CNM), while sightlines
with ${\rm N_{\sc HI}} \ge 2 \times 10^{20}$~per~cm$^{2}$ have low 
spin temperatures (median ${\rm T}_s \sim 240$~K), implying high CNM fractions. 
The threshold for CNM formation is likely to arise due to
inefficient self-shielding against ultraviolet photons at lower
\hi\ column densities. The threshold is similar to the defining
column density of a damped Lyman-$\alpha$ absorber; this indicates a
physical difference between damped and sub-damped Lyman-$\alpha$
systems, with the latter class of absorbers containing predominantly
warm gas.

\end{abstract}


\section{Introduction}
\label{sec:intro}

The diffuse interstellar medium (ISM) contains gas over a wide range
of densities, temperatures and ionization states. These can be broadly
sub-divided into the molecular phase (e.g. \citealt{snow06}), the 
neutral atomic phase (e.g. \citealt{kulkarni88}), and the warm and 
hot ionized phases (e.g. \citealt{haffner09}).
The neutral atomic medium (mostly neutral hydrogen, \hi) is further
usually sub-divided into ``cold'' and ``warm'' phases (the
``CNM'' and ``WNM'', respectively). Typical CNM temperatures and
densities are $\sim 40-200$~K and $\gtrsim 10$~cm$^{-3}$,
respectively, with corresponding WNM values of $\gtrsim 5000$~K and
$\sim 0.1-1$~cm$^{-3}$. This was originally an observational 
definition, to distinguish between phases producing strong narrow 
absorption lines towards background radio-loud quasars and
smooth broad emission lines \citep{clark65}. Later, this separation
into cold and warm phases was found to arise naturally in the context
of models in which the atomic and ionized phases are in pressure 
equilibrium.  Atomic gas at intermediate temperatures was found 
to be unstable to perturbations, and to migrate to either the cold 
or warm stable phases, given sufficient time to attain pressure 
equilibrium \citep{field69,wolfire95}.

Radio studies of the \hi-21cm transition have played a vital role
in understanding physical conditions in the neutral ISM. 
The \hi-21cm excitation temperature of a gas cloud (the ``spin temperature'', 
$\ts$) can be shown to depend on the local kinetic temperature ($\tk$; 
\citealt{field59,liszt01}), with $\ts \approx \tk$ in the CNM and 
$\ts \le \tk$ in the WNM \citep{liszt01}. Since the \hi-21cm absorption opacity 
for a fixed \hi\ column density varies inversely with $\ts$, while the emissivity
is independent of $\ts$ (in the low opacity limit), \hi-21cm absorption 
studies against background radio sources are primarily sensitive to the 
presence of CNM along the sightline, while \hi-21cm emission studies 
are sensitive to both warm and cold \hi. Indeed, the original evidence for 
two temperature phases in the neutral gas stemmed from a comparison between 
\hi-21cm absorption and emission spectra \citep{clark65}.

Physical conditions in the neutral gas are also known to depend on the
gas column density. At the low \hi\ column densities of the
inter-galactic medium, $\nhi < 10^{17}$~\cm, the gas is optically
thin to ionizing ultraviolet (UV) photons and is hence mostly ionized, 
giving rise to the so-called Lyman-$\alpha$ forest \citep{rauch98}.  
At higher \hi\ column densities, $10^{17}
\lesssim \nhi \lesssim 10^{20}$~\cm, the \hi\ is optically-thick to
ionizing photons at the Lyman-limit, and the core of the Lyman-$\alpha$
absorption line is saturated. Such ``Lyman-limit systems'' are partially 
ionized and typically arise for sightlines through the outskirts of galaxies
\citep{bergeron91}.  At still higher \hi\ column densities, 
$\nhi \ge 2 \times 10^{20}$~\cm, the Lyman-$\alpha$ line becomes
optically-thick in its naturally-broadened wings and acquires a
Lorentzian shape; such systems are called damped Lyman-$\alpha$
absorbers (DLAs; \citealt{wolfe05}). Damped absorption is ubiquitous 
on sightlines through galaxy disks and, at high redshifts, has long 
been used as the signature of the presence of an intervening galaxy along 
a quasar sightline \citep{wolfe86}. Finally, the molecular hydrogen fraction
shows a steep transition from very low values to greater than 5\% at $\nhi
\sim 5 \times 10^{20}$~\cm\ in the Galaxy \citep{savage77,gillmon06}). 
Neutral gas is likely to become predominantly molecular at higher \hi\ 
column densities, $\nhi \gtrsim 10^{22}$~\cm\ \citep{schaye01,krumholz09b}.

While it is well known that a threshold column density ($\nhi
\sim 5 \times 10^{20}$~\cm) is required to form the molecular phase, 
for self-shielding against UV photons in the ${\rm H}_2$ Lyman band
\citep{stecher67,hollenbach71,federman79}, the conditions for CNM
formation are less clear. In this {\it Letter}, we report results from
\hi-21cm absorption studies of a large sample of compact radio sources 
that indicate the presence of a similar, albeit lower, \hi\ column density 
threshold for CNM formation in the Milky Way.

\section{The sample}
\label{sec:sample}

We have used the Westerbork Synthesis Radio Telescope (WSRT, 23 sources), the 
Giant Metrewave Radio Telescope (GMRT, 10 sources) and the Australia Telescope 
Compact Array (ATCA, 2 sources) to carry out sensitive, high spectral resolution
($\sim 0.26-0.52$~km/s) \hi-21cm absorption spectroscopy towards 35~compact 
radio-loud quasars. Most of the target sources were selected to be B-array 
calibrators for the Very Large Array, and have angular sizes $\lesssim 5''$.  
Details of the sample selection, observations and data analysis are given in 
\citet{kanekar03b}, \citet{braun05}, Roy et al. (2011, in prep.) and Kanekar 
\& Braun (2011, in prep.). The final optical-depth spectra have root-mean-square 
(RMS) noise values of $\tau_{\rm RMS} \sim 0.0002 - 0.0013$ per 1~km/s channel, 
with a median RMS noise of $\sim 5 \times 10^{-4}$ per 1~km/s channel. These RMS 
noise values are at off-line channels, away from \hi-21cm emission that increases 
the system temperature, and hence, the RMS noise.  A careful bandpass calibration 
procedure was used to ensure a high spectral dynamic range and excellent sensitivity 
to wide absorption [see \citet{kanekar03b} and \citet{braun05} for details].
These are among the deepest \hi-21cm absorption spectra ever obtained 
(e.g. \citealt{dwaraka02,kanekar03b,begum10}) and constitute by far the largest 
sample of absorption spectra of this sensitivity. The use of interferometry also 
implies that the spectra are not contaminated by \hi-21cm emission within the 
beam, unlike the case with single-dish absorption spectra \citep{kanekar03b,heiles03}.

Galactic \hi-21cm absorption was detected against every source but
one, B0438-436. We used the Leiden-Argentine-Bonn (LAB) survey 
\citep{kalberla05}\footnote{http://www.astro.uni-bonn.de/\~webaiub/english/tools\_labsurvey.php}
to estimate the ``apparent'' \hi\ column density (uncorrected for self-absorption, i.e. 
$\nhi = 1.823 \times 10^{18} \times \int T_B {\rm dV}$~\cm\ \citep{wilson09}, where 
$T_B$ is the brightness temperature) at a location adjacent to each target.  The 
observational results are summarized in Table~\ref{tab:summary}, whose columns contain
(1)~the quasar name,
(2)~its Galactic latitude, 
(3)~the \hi\ column density (uncorrected for self-absorption), in units 
of $10^{20}$~\cm, from the LAB survey,
(4)~the integrated \hi-21cm optical depth $\int \tau {\rm dV}$, in km/s, and
(5)~the harmonic-mean spin temperature $\ts$ in K, estimated assuming 
the low optical depth limit ($\ts = \int T_B {\rm dV} / \int \tau {\rm dV}$).

Finally, for the one source with undetected \hi-21cm absorption (B0438-436),
we quote a $3\sigma$ upper limit on the integrated \hi-21cm optical depth, 
assuming a Gaussian line profile with a full width at half maximum of 20~km/s, 
and the corresponding $3\sigma$ lower limit on the spin temperature.

\setcounter{table}{0}
\begin{table}
\begin{center}
\caption{The full sample of 35 sources.
\label{tab:summary}}
\begin{tabular}{|c|c|c|c|c|}
\hline
  QSO        & Latitude & N$_{\rm HI}\:^a$         &   $\int \tau \rm{dV}\:^b$  &     $\ts$        \\
             &  $b$     & $\times 10^{20}$ \cm &        \kms\           &       K          \\
\hline
 B0023-263  & -84.17  &  $  1.641  \pm  0.022   $ &  $ 0.025  \pm  0.005  $ &  $  3546  \pm  694  $  \\  
 B0114-211  & -81.47  &  $  1.380  \pm  0.034   $ &  $ 0.135  \pm  0.005  $ &  $   559  \pm   26  $  \\  
 B0117-156  & -76.42  &  $  1.418  \pm  0.034   $ &  $ 0.031  \pm  0.004  $ &  $  2538  \pm  353  $  \\  
 B0134+329  & -28.72  &  $  4.272  \pm  0.029   $ &  $ 0.443  \pm  0.002  $ &  $   530  \pm    4  $  \\  
 B0202+149  & -44.04  &  $  4.809  \pm  0.027   $ &  $ 0.747  \pm  0.005  $ &  $   353  \pm    3  $  \\  
 B0237-233  & -65.13  &  $  2.078  \pm  0.038   $ &  $ 0.294  \pm  0.004  $ &  $   388  \pm    9  $  \\  
 B0316+162  & -33.60  &  $  9.431  \pm  0.034   $ &  $ 2.964  \pm  0.004  $ &  $   175  \pm    1  $  \\  
 B0316+413  & -13.26  &  $ 13.234  \pm  0.028   $ &  $ 1.941  \pm  0.003  $ &  $   374  \pm    1  $  \\  
 B0355+508  &  -1.60  &  $ 74.344  \pm  0.079   $ &  $ 45.820  \pm  1.120 $ &  $    89  \pm    2  $  \\  
 B0404+768  &  18.33  &  $ 10.879  \pm  0.032   $ &  $ 1.945  \pm  0.005  $ &  $   307  \pm    1  $  \\  
 B0407-658  & -40.88  &  $  3.363  \pm  0.018   $ &  $ 0.548  \pm  0.007  $ &  $   337  \pm    5  $  \\  
 B0429+415  &  -4.34  &  $ 37.067  \pm  0.041   $ &  $ 10.879  \pm  0.007 $ &  $   187  \pm    1  $  \\  
 B0438-436  & -41.56  &  $  1.380  \pm  0.027   $ &  $ < 0.020            $ &  $  > 3785 $  \\  
 B0518+165  & -11.34  &  $ 20.616  \pm  0.035   $ &  $ 6.241  \pm  0.007  $ &  $   181  \pm    1  $  \\  
 B0531+194  &  -7.11  &  $ 26.609  \pm  0.036   $ &  $ 4.062  \pm  0.005  $ &  $   359  \pm    1  $  \\  
 B0538+498  &  10.30  &  $ 19.542  \pm  0.029   $ &  $ 5.618  \pm  0.003  $ &  $   191  \pm    1  $  \\  
 B0831+557  &  36.56  &  $  4.469  \pm  0.032   $ &  $ 0.483  \pm  0.006  $ &  $   507  \pm    7  $  \\  
 B0834-196  &  12.57  &  $  7.125  \pm  0.036   $ &  $ 0.973  \pm  0.005  $ &  $   402  \pm    3  $  \\  
 B0906+430  &  42.84  &  $  1.251  \pm  0.036   $ &  $ 0.051  \pm  0.003  $ &  $  1342  \pm   89  $  \\  
 B1151-348  &  26.34  &  $  7.732  \pm  0.031   $ &  $ 0.714  \pm  0.005  $ &  $   594  \pm    5  $  \\  
 B1245-197  &  42.88  &  $  3.802  \pm  0.037   $ &  $ 0.158  \pm  0.005  $ &  $  1323  \pm   40  $  \\  
 B1323+321  &  81.05  &  $  1.260  \pm  0.034   $ &  $ 0.083  \pm  0.003  $ &  $   836  \pm   40  $  \\  
 B1328+254  &  80.99  &  $  1.065  \pm  0.036   $ &  $ 0.021  \pm  0.003  $ &  $  2729  \pm  411  $  \\  
 B1328+307  &  80.67  &  $  1.197  \pm  0.030   $ &  $ 0.072  \pm  0.002  $ &  $   916  \pm   32  $  \\  
 B1345+125  &  70.17  &  $  1.957  \pm  0.024   $ &  $ 0.305  \pm  0.005  $ &  $   352  \pm    7  $  \\  
 B1611+343  &  46.38  &  $  1.318  \pm  0.034   $ &  $ 0.019  \pm  0.003  $ &  $  3873  \pm  565  $  \\  
 B1641+399  &  40.95  &  $  1.044  \pm  0.032   $ &  $ 0.009  \pm  0.002  $ &  $  6498  \pm 1760  $  \\  
 B1814-637  & -20.76  &  $  6.416  \pm  0.018   $ &  $ 0.997  \pm  0.007  $ &  $   353  \pm    3  $  \\  
 B1827-360  & -29.34  &  $  8.163  \pm  0.018   $ &  $ 1.542  \pm  0.003  $ &  $   290  \pm    1  $  \\  
 B1921-293  & -11.78  &  $  7.312  \pm  0.017   $ &  $ 1.446  \pm  0.006  $ &  $   277  \pm    1  $  \\  
 B2050+364  & -48.84  &  $ 27.638  \pm  0.038   $ &  $ 3.024  \pm  0.010  $ &  $   501  \pm    2  $  \\  
 B2200+420  &  -5.12  &  $ 17.119  \pm  0.042   $ &  $ 3.567  \pm  0.016  $ &  $   263  \pm    1  $  \\  
 B2203-188  & -51.16  &  $  2.432  \pm  0.018   $ &  $ 0.248  \pm  0.004  $ &  $   537  \pm   10  $  \\  
 B2223-052  & -10.44  &  $  4.561  \pm  0.030   $ &  $ 1.034  \pm  0.003  $ &  $   242  \pm    2  $  \\  
 B2348+643  &   2.56  &  $ 70.522  \pm  0.051   $ &  $ 32.517  \pm  0.031 $ &  $   119  \pm    1  $  \\  
            &         &                        &                        &                   \\
\hline
\end{tabular}
\vskip 0.01in
Notes: $^a$~From the LAB emission survey. $^b$~From our \hi-21cm absorption survey.
\end{center}
\end{table}

\section{Results: An $\nhi$ threshold for CNM formation}
\label{sec:results}

\begin{figure*}[t!]
\epsfig{file=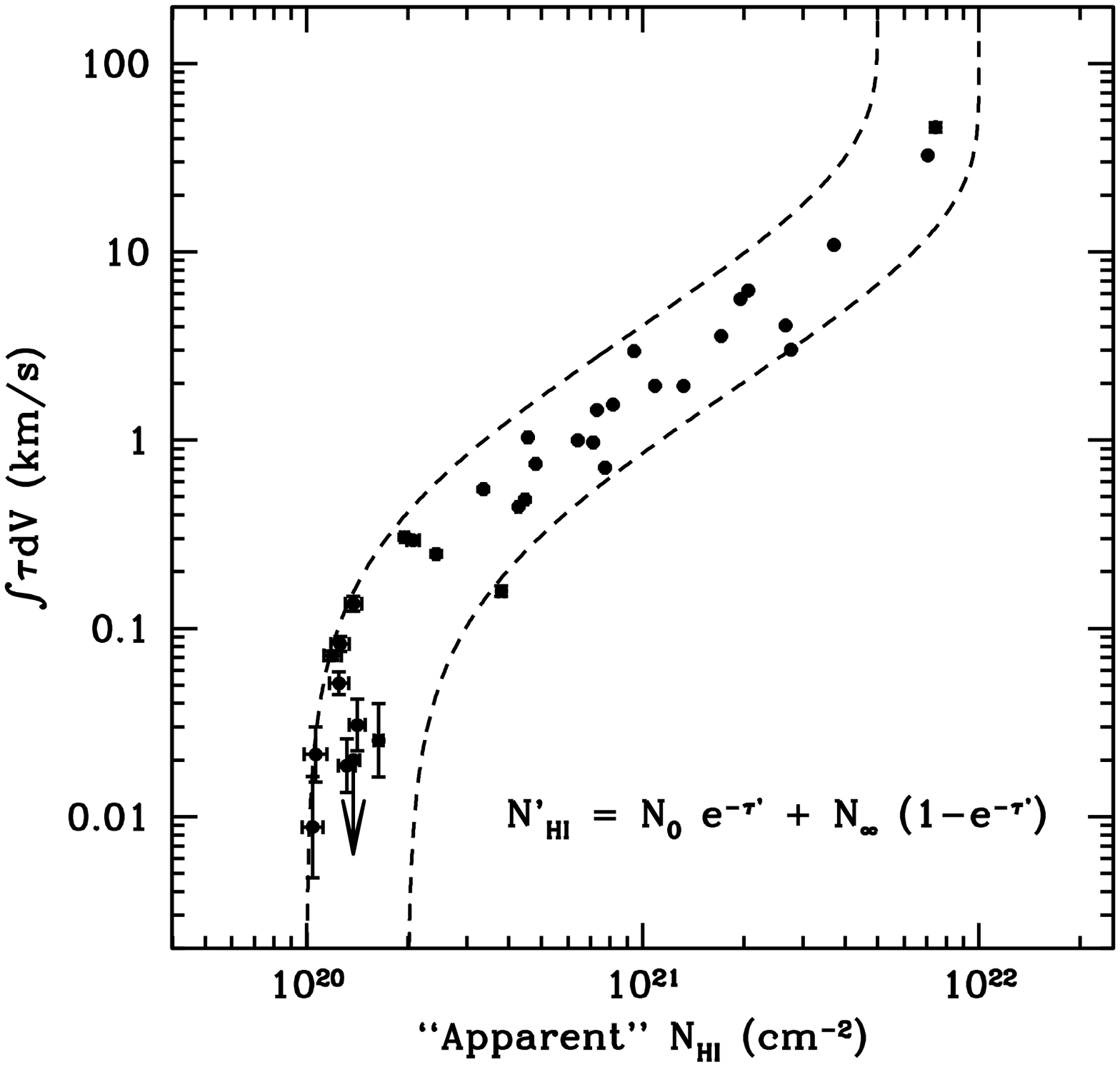,height=3.5truein,width=3.5truein}
\epsfig{file=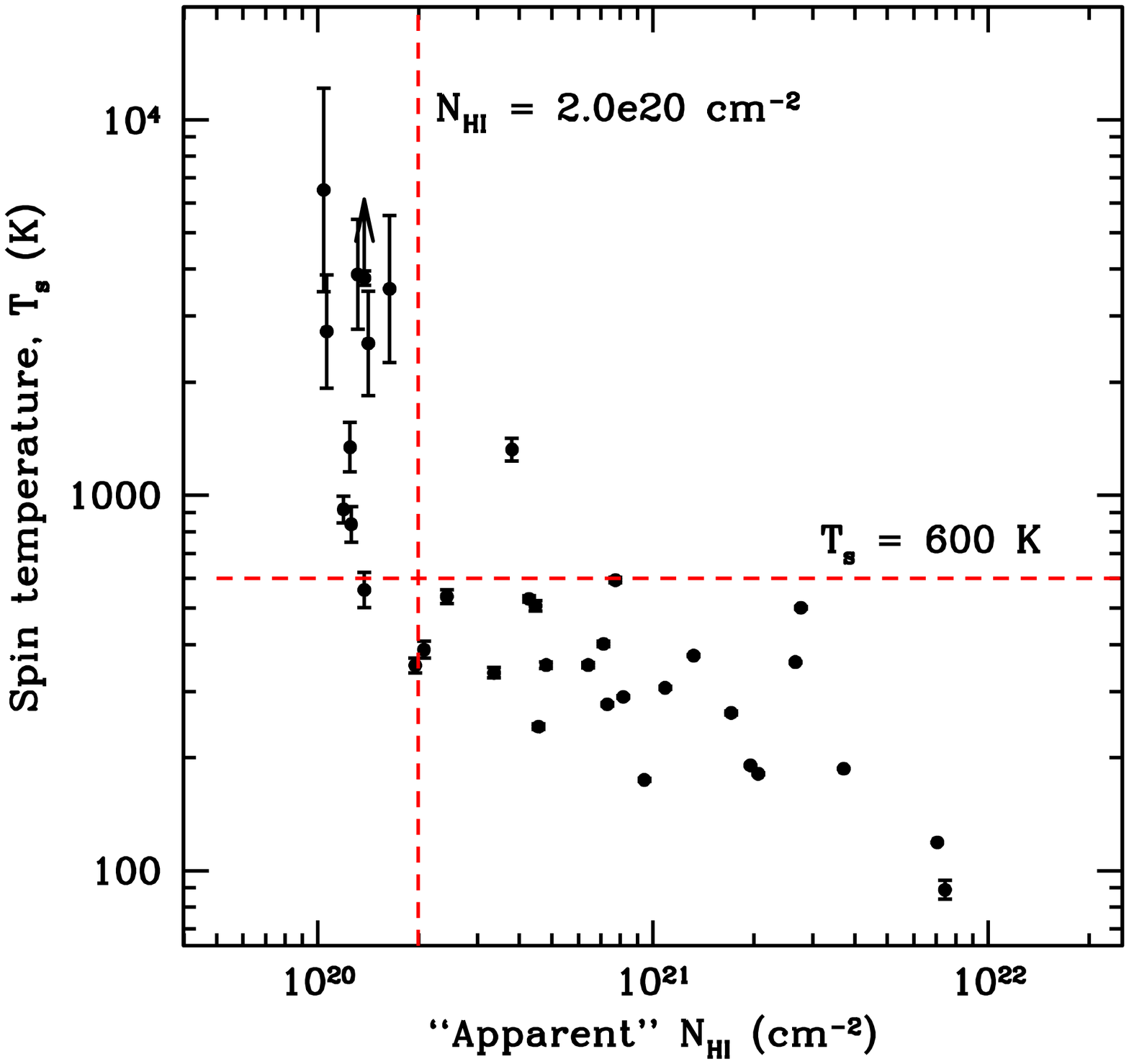,height=3.5truein,width=3.5truein}
\caption{[A]~Left panel: The integrated optical depth $\int \tau {\rm dV}$ (in km/s), 
from our \hi-21cm absorption survey, plotted against apparent \hi\ column density (\cm), 
from the LAB emission survey; the dashed curves show the relation in eqn.~\ref{eqn:nh} with $(log(N_0),
log(N_\infty), \Delta V)$ = (20.0, 21.7, 20) and (20.3, 22.0, 10).
[B]~Right panel: The column-density-weighted harmonic mean spin temperature (K)
plotted against apparent \hi\ column density (\cm). }
\label{fig:fig1}
\end{figure*}

The left panel of Figure~\ref{fig:fig1} shows the integrated \hi-21cm
optical depth plotted against \hi\ column density, on a logarithmic scale.  
It is clear from the figure that the integrated
\hi-21cm optical depth has a rough power-law dependence on $\nhi$ for $\nhi
\ge 2 \times 10^{20}$~\cm. However, there is a steep decline in $\int
\tau {\rm dV}$ for \hi\ column densities lower than this value. 
These properties were noted previously by \citet{braun92} for
individual spectral features (see their Figs.~3--6) rather than the
line-of-sight integral. A similar pattern is visible in the right
panel of the figure, which plots the spin temperature $\ts$ against
$\nhi$. 24 out of 25 sightlines with $\nhi \ge 2 \times
10^{20}$~\cm\ show $\ts < 600$~K, while all ten sightlines with $\nhi
< 2 \times 10^{20}$~\cm\ have $\ts > 550$~K. The median spin
temperature for sightlines with $\nhi \ge 2 \times 10^{20}$~\cm\ is
$\sim 240$~K, while that for sightlines with $\nhi < 2 \times
10^{20}$~K is $\sim 1800$~K. Thus, there appears to be a physical
difference between sightlines with \hi\ columns lower and higher than
the column density ${\rm N_{LIM}} = 2 \times 10^{20}$~\cm. Note that 
the use of apparent $\nhi$ only affects high-opacity sightlines,
and has no significant effect on our results.

We used a number of non-parametric two-sample tests (e.g. the Gehan test, 
the Peto-Prentice test, the logrank test, etc), generalized for
censored data, to determine whether the spin temperatures of
sightlines with $\nhi \ge \nlim$ and $\nhi < \nlim$ are drawn from the
same parent distribution. 
All tests used the {\sc ASURV} Rev.~1.2 package \citep{lavalley92}
which implements the methods of \citet{feigelson85}. The use
of multiple tests guards against any biases within a given test,
resulting from the relatively small sample size (10 systems with
$\nhi < \nlim$) and the presence of censored values \citep{feigelson85}. 
Errors on individual measurements were handled through a Monte-Carlo 
approach, using the measured values of $\ts$ and $\nhi$ (and the 
associated errors) for each sightline to generate $10^4$ sets of 35 
pairs of $\ts$ and $\nhi$ values. The statistical significance of each 
result (quoted below) is the average of values obtained in these 
$10^4$ trials for each test. The tests found clear evidence, 
with statistical significance between $4.2\sigma$ and $5.3\sigma$ 
that the spin temperatures for sightlines with $\nhi \ge \nlim$ and 
$\nhi < \nlim$ are drawn from different parent populations. Following 
\citet{feigelson85}, our final result is based on the Peto-Prentice test, 
as this has been found to give the best results for very different sample sizes
\citep{latta81}. The two $\ts$ sub-samples (with ``low'' and ``high'' 
$\nhi$) are then found to be drawn from different parent distributions at 
$5.3 \sigma$ significance; the probability of this arising by chance 
is $\sim 2 \times 10^{-7}$.

It thus appears that sightlines with $\nhi < \nlim = 2 \times 10^{20}$~\cm\ 
have systematically higher spin temperatures than sightlines with $\nhi > \nlim$.
The spin temperature measured here is the column-density-weighted 
harmonic mean of the spin temperatures of different ``phases'' along 
the line of sight (e.g. \citealt{kulkarni88}).  For example, a sightline 
with $\nhi$ equally divided between CNM and WNM, with $\ts = 100$~K and $\ts =
8000$~K respectively, would yield an average $\ts \sim 200$~K, while
one with 90\% of gas with $\ts = 8000$~K and 10\% with $\ts = 100$~K
would yield an average $\ts = 900$~K.  In other words, a high $\ts$
indicates the presence of a smaller fraction of the cold phase of \hi. 
Thus, the fact that average spin temperatures are significantly higher 
on sightlines with low \hi\ column densities, $\nhi < 2 \times 10^{20}$~\cm, 
indicates that such low-$\nhi$ sightlines contain low CNM fractions, 
far smaller than those on sightlines with $\nhi \ge 2 \times 10^{20}$~\cm.

It is clear from Fig.~\ref{fig:fig1}[A] that the integrated \hi-21cm
optical depth drops sharply at $\nhi = \nlim$, due to the decline in
the CNM fraction at low \hi\ column densities. This can be accounted
for by a simple physical model in which a minimum ``shielding''
\hi\ column density of WNM is needed for the formation of the cold
phase in an \hi\ cloud. Assuming that the \hi-21cm optical depth in
the WNM is negligible compared to that in the CNM, the equation
of radiative transfer for a ``sandwich'' geometry can be written as
\begin{equation}
T_B{\rm(V)}  =  \frac{T_w\tau_w}{2} + T_c (1-e^{-\tau_c}) + \frac{T_w\tau_w}{2} e^{-\tau_c}
\end{equation}
where the explicit velocity dependence of $\tau_W$ and $\tau_C$ has been
omitted for clarity and the integrated \hi\ emission profile becomes,
\begin{equation}
\int T_B {\rm dV} = \int T_w\tau_w e^{-\tau_c} {\rm dV} + \int \bigg(T_c + \frac{T_w\tau_w}{2}\bigg)
(1-e^{-\tau_c}) {\rm dV}
\end{equation}
or,
\begin{equation}
N'_{HI} = N_{0} e^{-\tau_c'} + N_{\infty} (1-e^{-\tau_c'}) 
\label{eqn:nh}
\end{equation}
for a measured ``apparent'' column density, $N'_{\rm HI}$ (assuming
negligible self-opacity), a threshold column density (where $\tau_c
\rightarrow 0$), $N_0 \sim T_w \tau_w \Delta V$, a saturation column 
density (where $\tau_c
\rightarrow \infty$), $N_\infty \sim (T_c+T_w \tau_w / 2)\Delta V$ and an
effective opacity, $\tau_c'$. The effective opacity is related to the
measured integrated opacity by the effective linewidth, $\Delta V$, as
$\int \tau {\rm dV} = \tau_c' \Delta V$.

The upper and lower dashed curves in Fig.~\ref{fig:fig1}[A] show the
above expression for (${\rm N_{0}} = 10^{20}$~\cm, ${\rm N_{\infty}} = 
5.0 \times 10^{21}$~\cm, $\Delta V = 20$~km/s) and 
(${\rm N_{0}} = 2 \times 10^{20}$~\cm, ${\rm N_{\infty}} = 10^{22}$~\cm, 
$\Delta V = 10$~km/s), respectively.  Note that the effective linewidth only 
shifts the curves up and down in the figure. The fact that none of the data 
points of Fig.~\ref{fig:fig1}[A] lie to the left of the upper curve indicates 
that a minimum {\it shielding} \hi\ column of $[N_{0}/2] \sim 5 \times 10^{19}$~\cm\ 
is needed for the formation of the cold neutral medium. Note that, in the above 
``sandwich'' geometry, a total WNM column density of $\gtrsim 10^{20}$~\cm\ is needed 
before any CNM can be formed.

\section{Discussion}
\label{sec:discuss}

\begin{figure}[t!]
\epsfig{file=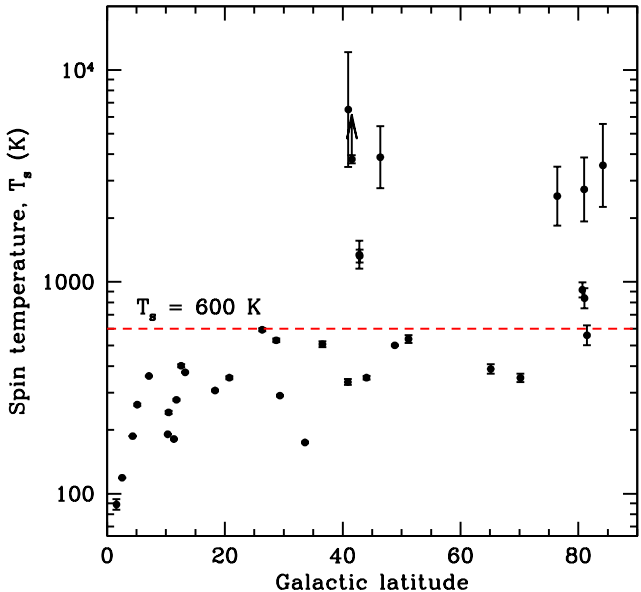,height=3.5truein,width=3.5truein}
\caption{The column-density-weighted harmonic mean spin temperature (K) plotted against 
Galactic latitude. 
 }
\label{fig:fig2}
\end{figure}

Most theoretical models of the diffuse interstellar medium treat it as
a multi-phase medium, consisting of neutral and ionized phases in
pressure equilibrium (e.g. \citealt{field69,mckee77,wolfire95}). In
the McKee-Ostriker model, physical conditions in the ISM are regulated
by supernova explosions, whose blast waves sweep up gas in the ISM
into shells, leaving large cavities full of hot ionized gas
\citep{mckee77}. Cold atomic gas is then formed by the rapid cooling
of the swept-up and shocked gas. Soft X-rays from neighbouring
hot ionized gas and stellar UV photons penetrate into
the CNM and heat and partially ionize it, producing envelopes of warm
neutral and warm ionized gas (WIM). Thus, a typical ISM ``cloud'' [see Fig.~1 of
  \citet{mckee77}] is expected to form at the edges of supernova
remnants and super-bubbles containing the hot ionized medium (HIM). Such a
cloud consists of a CNM core surrounded by a WNM envelope and a
further WIM shell, all of which are in pressure equilibrium. 
The CNM core is almost entirely neutral, the WNM and WIM are partially 
ionized, and the HIM is almost entirely ionized \citep{mckee77}.

A threshold \hi\ column density for CNM formation in an ISM cloud
arises quite naturally due to the need for self-shielding against
ionizing ultraviolet (and soft X-ray) photons (see also \citealt{schaye04}). At low \hi\ columns, UV
photons can penetrate all the way into a cloud core and heat (and
partly ionize) the \hi. Only clouds that self-shield against the
penetration of these high-energy photons can retain a stable CNM
core. Self-shielding against UV photons is significant once the
optical depth at the Lyman limit crosses unity, i.e. for $\nhi \ge
10^{17}$~\cm, and becomes more and more efficient with increasing
\hi\ column density. Our results indicate that self-shielding only
becomes entirely efficient at excluding UV photons from the interior
of \hi\ clouds at a {\it total} \hi\ column density of $\nhi \sim 2 \times
10^{20}$~\cm.  Below this threshold, sufficient UV photons penetrate
into the cloud interior to hinder the survival of the cold phase. The
simple ``sandwich'' model overlaid on Fig.~\ref{fig:fig1}[A] suggests
that a shielding column of $\sim (0.5 - 0.75) \times 10^{20}$~\cm\ {\it 
on each exposed surface of a cloud} is sufficient to exclude UV photons,
with the rest of the \hi\ then free to cool to lower temperatures.

It is also possible that sightlines with low $\nhi$ are sampling gas
at higher distances from the Galactic plane, with lower metallicity
and pressure.  Two-phase models that incorporate detailed balancing of
heating and cooling rates obtain lower CNM fractions at lower
pressures and metallicities \citep{wolfire95}.  Observationally,
\citet{kanekar09c} have found evidence for an anti-correlation between
spin temperature and metallicity in high-$z$ DLAs, with low-metallicity 
DLAs having higher spin temperatures, probably because the paucity of metals 
yields fewer cooling routes \citep{kanekar01a}.  If low-$\nhi$ sightlines
contain ``clouds'' with systematically lower metallicities (and/or pressures), 
these could have lower cooling rates and hence, lower CNM fractions. 
Unfortunately, we do not have estimates of either pressure or metallicity 
along our sightlines and hence cannot test this possibility. We note that 
all sightlines with high spin temperatures are at intermediate or high Galactic 
latitudes, $b > 40^\circ$ (see Fig.~\ref{fig:fig2}), suggesting larger distances
from the plane.  Conversely, the figure also shows that low spin
temperatures ($\lesssim 400$~K) are obtained even for sightlines at
high Galactic latitudes ($b \sim 40-70^\circ$). Measurements 
of the metallicity and pressure along the low-$\nhi$ sightlines, through
UV spectroscopy, would be of much interest.

There have been earlier suggestions, based on semi-analytical or 
numerical treatments of self-shielding, that \hi\ clouds are predominantly 
neutral for $\nhi \gtrsim 10^{20}$~\cm\ (e.g. \citealt{viegas95,wolfe05}). 
However, this is the first direct evidence for a change in physical conditions 
in \hi\ clouds at this column density. Note that this is a factor of several
lower than the known threshold of $\nhi \sim 5 \times 10^{20}$~\cm\ for 
the formation of molecular hydrogen in Galactic clouds, set by the requirement 
of self-shielding against UV photons at wavelengths in the H$_2$ Lyman band 
\citep{savage77}. Thus, there appear to be three column densities at which phase
transitions occur in ISM clouds, at $\nhi \sim 2 \times
10^{20}$~\cm\ resulting in the formation of cold \hi, at $\nhi \sim 5
\times 10^{20}$~\cm\ resulting in the formation of molecular hydrogen,
and finally, at $\nhi > 10^{22}$~\cm, when most of the atomic gas
is converted into the molecular phase.

In this context, it is vital to appreciate that the abscissa of
Fig.~\ref{fig:fig1} refers to {\it apparent\ } \hi\ column density under
the assumption of negligible self-opacity of the \hi\ profile. The
saturation column density, N$_\infty$, is an {\it observational} upper limit
to $\int T_B dV$ and does not represent a physical limit on $\nhi$. In
fact, the widespread occurrence of the \hi\ self-absorption phenomenon
within the Galaxy \citep{gibson05} and the detailed modeling of high
resolution extragalactic \hi\ spectra suggests that self-opacity is a
common occurrence \citep{braun09} which can disguise neutral column 
densities that reach $\nhi \sim 10^{23}$~\cm.

The original definition of a DLA as an absorber with $\nhi \ge 2
\times 10^{20}$~\cm\ was an observational one, based on the
requirement that the damping wings of the Lyman-$\alpha$ line be
detectable in low-resolution optical spectra of moderate sensitivity
\citep{wolfe86}.  With today's 10-m class optical telescopes, it is
easy to detect the damping wings at significantly lower \hi\ column
densities, $\sim 10^{19}$~\cm (e.g. \citealt{peroux03}); such systems
are referred to as ``sub-DLAs''. There has been much debate in the
literature on whether or not DLAs and sub-DLAs should be treated as a
single class of absorber and on their relative importance in
contributing to the cosmic budget of neutral hydrogen and metals
(e.g. \citealt{peroux03,wolfe05,kulkarni10}).  \citet{wolfe05} argue
that DLAs and sub-DLAs are physically different, claiming that most of 
the \hi\ in sub-DLAs is ionized and at high temperature, while
that in DLAs is mostly neutral; this is based on numerical estimates
of self-shielding in DLAs and sub-DLAs against the high-$z$ UV
background \citep{viegas95}. Our detection of an \hi\ column density
threshold for CNM formation that matches the defining DLA column
density indicates that DLAs and sub-DLAs are indeed physically
different types of absorbers, with sub-DLAs likely to have significantly 
lower CNM fractions than DLAs, at a given metallicity.

In summary, we report the discovery of a threshold \hi\ column density, 
$\nhi \sim 2 \times 10^{20}$~\cm, for cold gas formation in \hi\ 
clouds in the interstellar medium. Above this threshold, the majority 
of Galactic sightlines have low spin temperatures, $\ts \lesssim 500$~K, 
with a median value of $\sim 240$~K. Below this threshold, typical sightlines
have far higher spin temperatures, $> 600$~K, with a median value of 
$\sim 1800$~K. The threshold for CNM formation appears to arise naturally 
due to the need for self-shielding against ultraviolet photons, which
penetrate into the cloud at lower \hi\ columns and heat and ionize the 
\hi, inhibiting the formation of the cold neutral medium.

\acknowledgments
We thank the staff of the GMRT and WSRT for help during the observations. The 
GMRT is run by the National Centre for Radio Astrophysics of the Tata Institute 
of Fundamental Research. The WSRT is operated by ASTRON with support from the 
Netherlands Foundation for Scientific Research (NWO). NK acknowledges support 
from the Department of Science and Technology through a Ramanujan Fellowship.
The National Radio Astronomy Observatory is a facility of the National Science 
Foundation operated under cooperative agreement by Associated Universities, Inc.

\bibliographystyle{apj}

\end{document}